\documentclass[journal=jacsat,manuscript=article]{achemso}

\usepackage[version=3]{mhchem} % Formula subscripts using \ce{}
\usepackage[utf8]{inputenc}
\usepackage[T1]{fontenc}
\usepackage{siunitx}
\usepackage{xcolor} 
\newcommand{\authormark}[1]{\textsuperscript{#1}}

\author{Evelyn Díaz-Escobar\authormark{1}}
\affiliation{Nanophotonics Technology Center, Universitat Politècnica de València, Camino de Vera s/n, 46022, Valencia, Spain}
\author{Ángela I. Barreda\authormark{1}}
\affiliation{Friedrich Schiller University Jena, Institute of Solid State Physics, Max-Wien-Platz 1, 07743 Jena, Germany}
\alsoaffiliation{Nanophotonics Technology Center, Universitat Politècnica de València, Camino de Vera s/n, 46022, Valencia, Spain}
\author{Amadeu Griol}
\affiliation{Nanophotonics Technology Center, Universitat Politècnica de València, Camino de Vera s/n, 46022, Valencia, Spain}
\author{Alejandro Martínez}
\affiliation{Nanophotonics Technology Center, Universitat Politècnica de València, Camino de Vera s/n, 46022, Valencia, Spain}
\email{amartinez@ntc.upv.es}

\title
  {Experimental observation of higher-order anapoles in individual silicon disks under in-plane illumination}

\abbreviations{IR,NMR,UV}

\begin{document} 

\authormark{1}{these authors contributed equally to this work}


\begin{abstract}
 Anapole states - characterized by a strong suppression of far-field scattering - naturally arise in high-index nanoparticles as a result of the interference between certain multipolar moments. Recently, the first-order electric anapole, resulting from the interference between the electric and toroidal dipoles, was characterized under in-plane illumination as required in on-chip photonics. Here, we go a step further and report on the observation of higher-order (magnetic and second-order electric) anapole states in individual silicon disks under in-plane illumination. To do so, we increase the disk dimensions (radius and thickness) so that such anapoles occur at telecom wavelengths. Experiments show dips in the far-field scattering perpendicular to the disk plane at the expected wavelengths and the selected polarizations, which we interpret as a signature of high-order anapoles. Some differences between normal and in-plane excitation are discussed, in particular the non-cancellation of the sum of the Cartesian electric and toroidal moments for in-plane incidence. Our results pave the way towards the use of different anapole states in photonic integrated circuits, either on silicon or other high-index dielectric materials.
\end{abstract}

\section{Introduction}
Interference between different radiative moments supported by high-index dielectric particles can eventually lead to scattering cancellation in the far-field, resulting in the so-called anapole states \cite{Miroshnichenko2015,Wang2016,Baryshnikova2019}. Interestingly, such non-radiative states are accompanied by a strong field concentration inside the nanoparticle\cite{Yang2018,Baryshnikova2019,Savinov2019}, a feature that has been used to boost light-matter interaction and enhance nonlinear effects such as harmonic generation \cite{Grinblat2017,Timofeeva2018} or Raman scattering \cite{Baranov2018}. One of the most interesting nanostructures to study anapole states is a high-index disk, since its radius can be defined lithographically to tune the anapole mode at a given frequency \cite{Miroshnichenko2015,Yang2018,Diaz2021}. Most studies of anapole states in high-index dielectric disks have addressed electric anapoles that arise from the destructive interference between the Cartesian electric and toroidal dipoles, which can be excited upon normal illumination \cite{Miroshnichenko2015, review_1, review_2}. Indeed, multiple higher-order electric anapoles arise when the wavelength-to-radius wavelength is increased \cite{Zenin2017,Yang2018,Tian2019}. But anapoles can also have a magnetic character when involving the magnetic and toroidal dipoles \cite{Luk2017,Lamprianidis2018,Baryshnikova2019}. The existence of magnetic anapoles in dielectric disks under normal illumination requires larger aspect ratios, which has led to the recent observation of these states at microwave frequencies \cite{Kapitanova2020}.

In Ref. \cite{Diaz2021}, we reported that the first electric anapole state can also be observed when the disk is illuminated in-plane using an integrated waveguide propagating the transverse-electric (TE) mode. Interestingly, we found that the electric anapole, characterized in the far-field by a huge dip in the top scattering, takes place at a different wavelength than the energy maximum in the disk, which was attributed to retardation effects in the disk. In principle, the same approach could be used to observe higher-order anapoles as long as the disk dimensions are properly tailored to support such states, and the radiation of the multipoles that contribute to the anapole formation takes place in the vertical (or out-of-plane) direction, which is the one accessible in experiments (see details below).    

In this work, we analyze both numerically and experimentally the in-plane (on-chip) realization of the second-order electric and the magnetic anapole states in an individual silicon disk built in a photonic integrated circuit (PIC). The second-order electric anapole can be observed following the same procedure as in \cite{Diaz2021} - this is, using the TE waveguide mode to excite the transverse electric dipole - but making the disk bigger. To this end, we increase the silicon thickness from the standard 220 nm thickness up to 350 nm, as well as, the disk radius, to place the anapole at the telecom wavelengths window.  To observe signatures of magnetic anapoles, we have to address excitation of the magnetic field in the disk along the transversal direction, since an ideal vertical magnetic (which could be excited using the TE waveguide mode) dipole does not radiate out-of-plane. This is achieved by using the transverse magnetic (TM) mode of the driving silicon waveguide. We observe strong dips - whose central wavelengths depend on the disk radius - in the far-field scattering spectra for the relevant polarizations,  which we consider a signature of the excitation of the targeted second-order electric and magnetic anapoles \cite{Kapitanova2020}. We also observe important differences between normal and in-plan excitation, remarkably for the second-order electric anapole, which we discuss in the text. Our results confirm the potential of this integrated approach to excite complex multipolar states in isolated high-index disks, opening new avenues to use anapole states in PICs. 

\section{Description of the system}
\begin{figure}[h!]
      \centering
      \includegraphics[width=\textwidth]{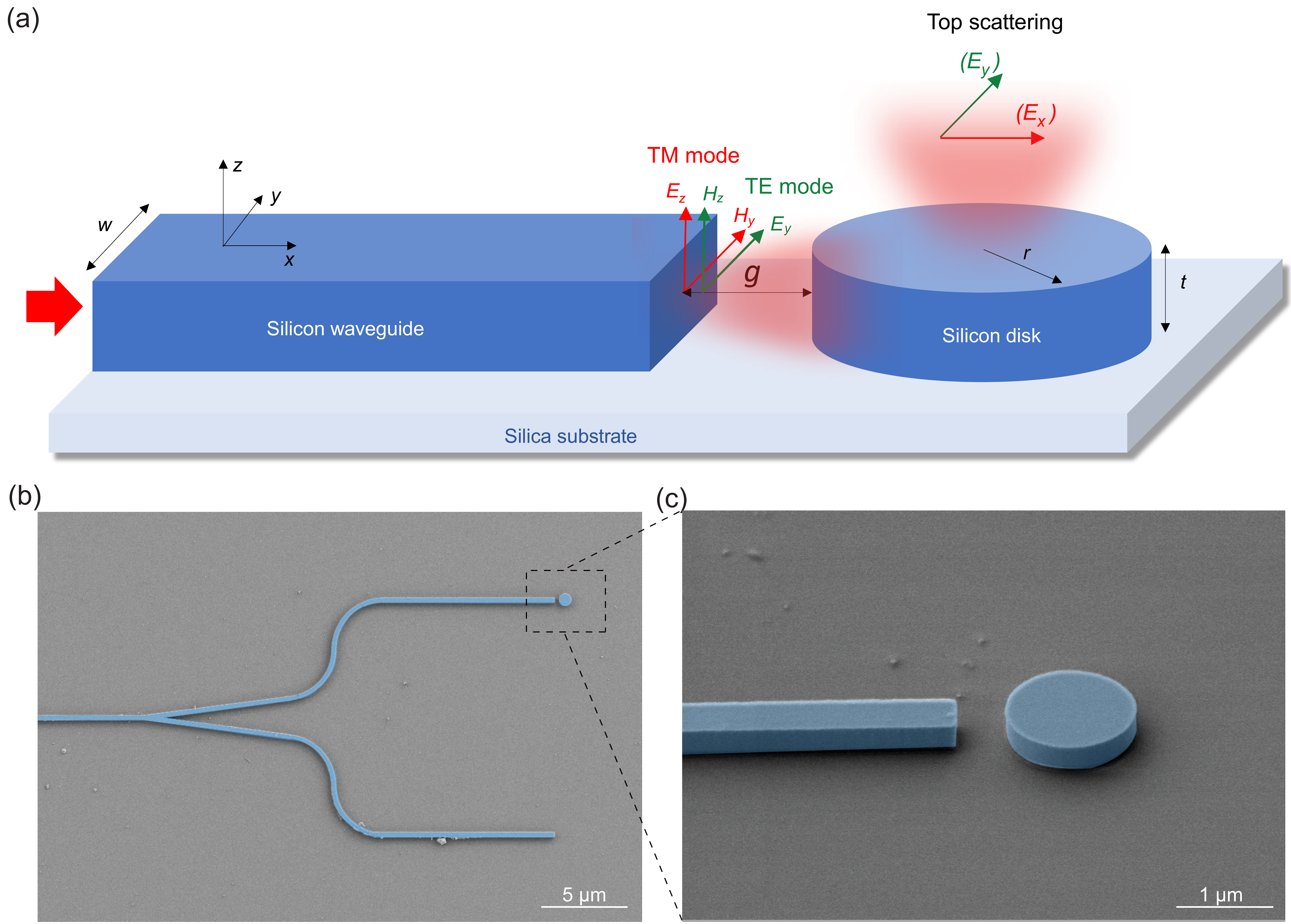}
      \caption{(\textbf{a}) Sketch of the in-plane excitation of a silicon disk (radius $r$, thickness $t$) from a waveguide (width $w$) end propagating either the TE or the TM mode. The main electric field components for each mode are depicted. The waveguide end and the disk are spaced by a gap $g$. The excited electric (magnetic) state scatters light mostly polarized along the $y$ ($x$) direction. (\textbf{b}) SEM image of one of the fabricated circuits, showing the input waveguide, a Y-splitter, and two waveguides, both ended in an abrupt termination, and one of them having a disk. (\textbf{c}) SEM image showing in detail the disk and the waveguide termination. Nominal disk radius $r = {550}$ nm, thickness $t = {350}$ nm, waveguide width $w = {450}$ nm and gap $g = {500}$ nm.}
      \label{fig:fig1}
\end{figure}

Figure \ref{fig:fig1}(a) displays a sketch of the structure used to observe the higher-order anapole states: an $x$-axis-oriented silicon waveguide with rectangular cross-section can propagate either its fundamental TE or TM mode characterized by its main field components $E_{\mathrm{z}}$ and $H_{\mathrm{y}}$ or $E_{\mathrm{y}}$ and $H_{\mathrm{z}}$. The waveguide is abruptly terminated, and the output fields illuminating the disk (separated from the waveguide by a gap $g$) are essentially the same as in the waveguide \cite{Espinosa-Soria2016}. Notice that guided modes are characterized by a main transverse component but they also carry longitudinal field components \cite{Espinosa-Soria2016_2} that are disregarded in this work for the sake of simplicity. Thus, the field exiting the waveguide can excite either the electric dipole or the magnetic dipole along the $y$-axis depending on the propagating waveguide mode (TE or TM, respectively). Ideally, an $y$-axis oriented electric (magnetic) dipole will radiate in a doughnut-shape pattern so that the electric far-field in the out-of-plane direction will be polarized along $y$ ($x$)   \cite{Savinov2019}. Therefore, in analogy to the case of the electric anapole \cite{Diaz2021}, detecting a reduction of the far-field scattering when measuring either the $E_{\mathrm{y}}$ or the $E_{\mathrm{x}}$ component could be interpreted as a signature of a second-order electric or a magnetic anapole state, respectively. Notably, the use of the TM mode also allows to excite the fundamental $z$-axis electric dipole but, in contrast to \cite{Kapitanova2020}, this component will not radiate in the vertical direction, which makes easier to detect the magnetic anapole. 

\section{Numerical simulations}

\begin{figure}[h!]
      \centering
      \includegraphics[width=0.7\textwidth]{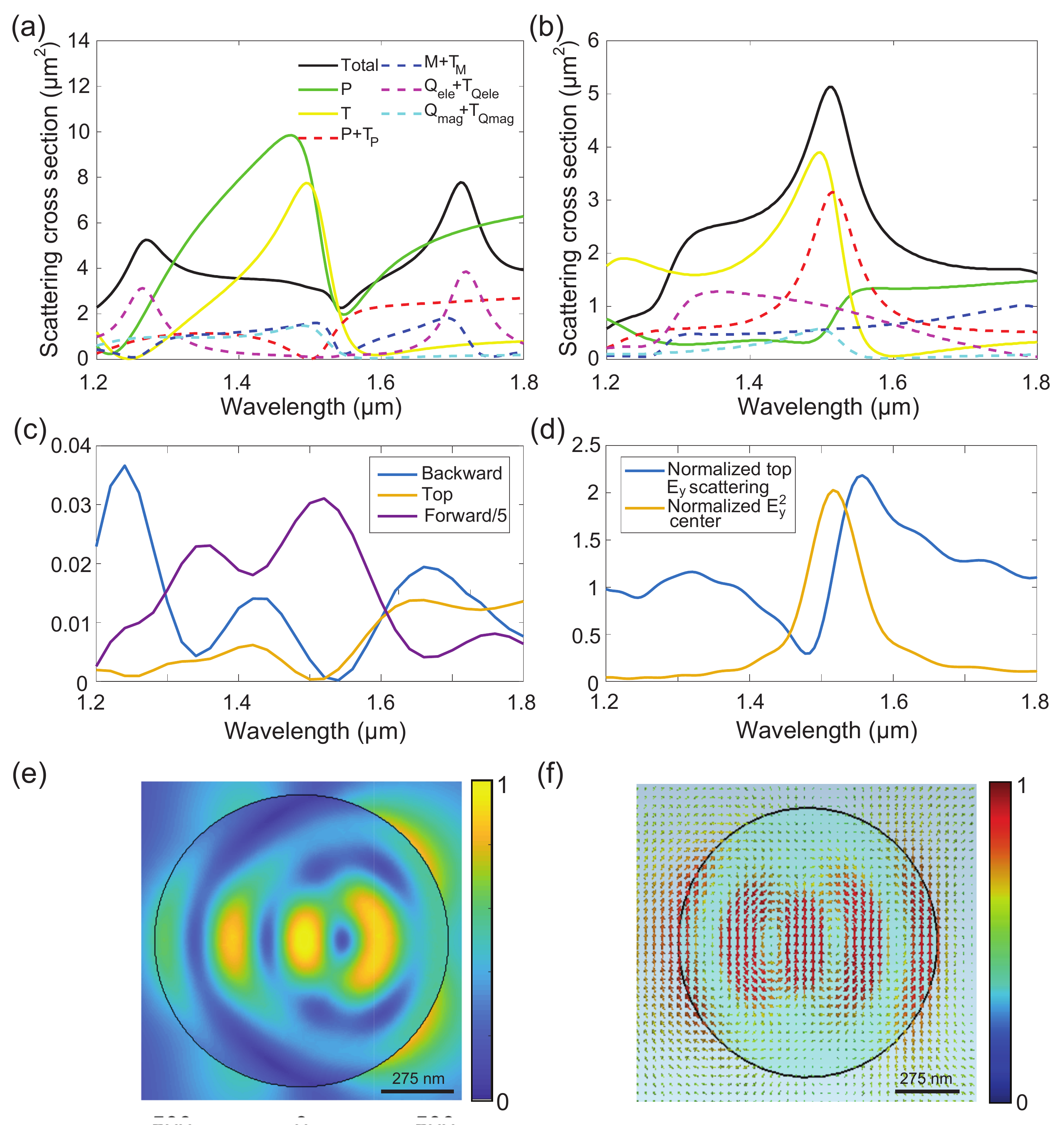}
      \caption{\textbf{Scattering and near-field simulations of a silicon  disk with $t = \SI{350}{nm}$ and  $r = \SI{550}{nm}$ to observe the second electric anapole}. Spectral contribution under (\textbf{a}) normal and (\textbf{b}) lateral illumination of the dipolar electric (\textbf{p}) in red and magnetic (\textbf{m}) in blue, and quadrupolar electric ($Q_{\mathrm{ele}}$) in purple and magnetic ($Q_{\mathrm{mag}}$) in cyan moments, together with their respective toroidal moments, represented by $T$, to the scattering cross-section. Also, the individual contributions of the electric dipole (green), and its respective toroidal moment (yellow) to the scattering are represented; (\textbf{c}) Scattering ($E_{\mathrm{y}}$ component) of the disk illuminated in free space (FS) by a plane wave with the electric field parallel to the disk axis in the forward, backward and out-of-plane (top) directions. The top scattering shows a minimum around the wavelength for which the second electric anapole is predicted for normal illumination; (\textbf{d}) Normalized top-scattering of the disk (blue curve) when illuminated by the light exiting from a waveguide propagating the TE mode. The yellow curve shows the normalized $|E_{\mathrm{y}}|^2$ component at the center of the disk;  (\textbf{e}) Near-field pattern of the norm of the electric field at the anapole state wavelength, normalized to its maximum value, at $\lambda=1510$ nm; (\textbf{f}) Electric field lines at at $\lambda=1510$ nm under waveguide illumination. Details about the simulations are given in the Supplementary Information.}
      \label{fig:fig2}
\end{figure}

We start our study by calculating the scattering properties of a single silicon disk surrounded by air. The effect of the substrate in the experiments will mainly consist on a red-shift of the whole optical response \cite{Diaz2021}. We fix the disk thickness to $t = {350}$ nm, which is thicker than the ${220}$ nm value typical in silicon-on-insulator photonics but still available in commercial wafers. We choose a radius $r={550}$ nm and perform numerical simulations of the disk response upon normal and lateral illumination with the electric field along the $x$ direction. Therefore, the lateral illumination mimics the scheme in Fig. \ref{fig:fig1} for the TE waveguide mode. The contributions to the scattering cross-section of the main multipole moments (Cartesian electric \textbf{p}, magnetic \textbf{m}, quadrupolar electric $Q_{\mathrm{ele}}$ and quadrupolar magnetic $Q_{\mathrm{mag}}$, together with their respective toroidal moments (\textbf{T}) are shown in Figs. \ref{fig:fig2}(a) and (b) for normal and lateral illumination, respectively. It can be seen (Fig. \ref{fig:fig2}(a)) that the second electric anapole condition - this is, $P+T_{P}=0$ - is satisfied at a wavelength of $\SI{1475}{nm}$ under normal incidence.  However, this condition is not met for lateral incidence (Fig. \ref{fig:fig2}(b)). So, if for the first electric anapole, we identified a major difference between in-plane and normal incidence (the decoupling of the scattering minimum and energy maximum in the disk \cite{Diaz2021}), for the second electric anapole, the difference is even more noticeable, and there is neither a scattering minimum nor a dip in $P+T_{P}$ at the expected (from normal incidence simulations) anapole wavelength. 

This behavior can be also attributed to retardation effects: the disk diameter is comparable to the wavelength which may result in different scattering properties depending on the illumination direction. However, in on-chip photonics, the key requirement is to ensure minimum out-of-plane scattering (so that light remains in the PIC) whilst in-plane scattering can occur and even used to build - for instance - chains of disks for energy transport \cite{Mazzone2017,Huang2021}. To verify if the scattering is reduced in the normal direction (parallel to the disk axis), we performed finite domain time domain (FDTD) simulations and calculated the scattered $E_{\mathrm{y}}$ field along different directions. As shown in Fig. \ref{fig:fig2}(c) there is a dip in the scattering response in the wavelength region where the electric anapole is predicted from normal excitation. This result also keeps when the disk is illuminated from the waveguide propagating the TE mode. As shown in Fig. \ref{fig:fig2}(d), the normalized top scattering (calculated as the ratio between the scattering with and without disk to remove the waveguide termination effect \cite{Diaz2021}) also displays a strong reduction in the expected wavelength region. Notice also that the electric field inside the disk also displays a maximum at a wavelength slightly red-shifted compared to the scattering. This means that the decoupling of the minimum scattering condition and the maximum-energy enhancement (we assume that the maximum energy takes place when the electric field in the disk center is maximized, as we observed in Ref.\cite{Diaz2021}) also takes place for the second electric anapole. 

Figures \ref{fig:fig2}(e) and (f) show the electric near-field at $\lambda=1510$ nm (the wavelength showing the top scattering minimum in Fig. \ref{fig:fig2}(c)) on the symmetry plane of the disk under left-side lateral illumination either from free space or from a waveguide, respectively (see details in the Supplementary Information). We observe field patterns similar to those occurring under normal illumination \cite{Yang2018} in the sense of having five field maxima separated by four nodes (forming four vortices of the electric field, as shown by the arrows depicted in Fig. \ref{fig:fig2}(f)). The asymmetry in the near-field is due to the symmetry-breaking illumination conditions\cite{Diaz2021}.

\begin{figure}[htbp]
      \centering
      \includegraphics[width=0.7\textwidth]{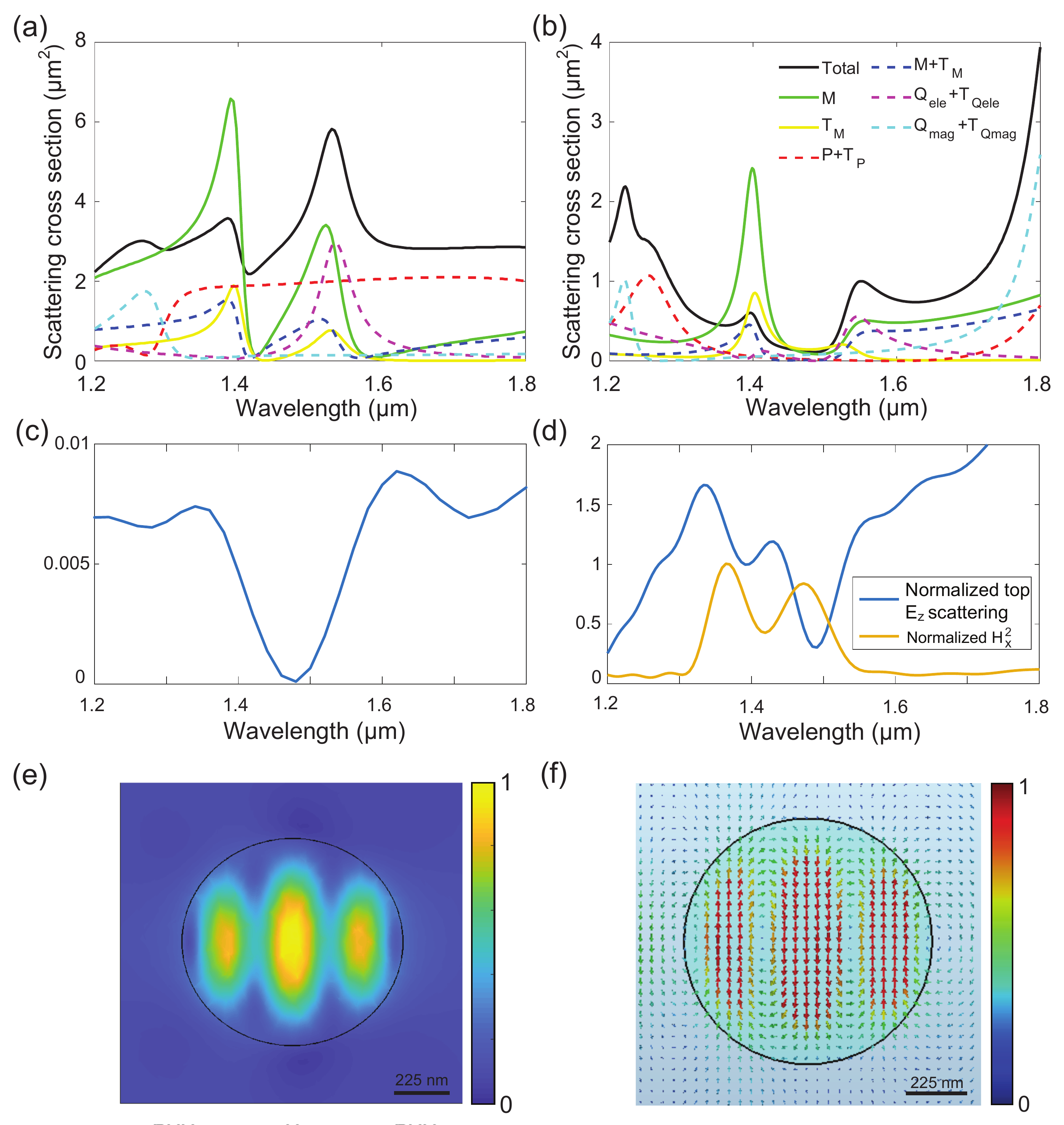}
      \caption{\textbf{Scattering and near-field simulations of a silicon  disk with $t = \SI{350}{nm}$ and  $r = \SI{450}{nm}$} to observe a magnetic anapole. Panels (\textbf{a}) and (\textbf{b}) are as in Fig. \ref{fig:fig2} but for the new illumination; (\textbf{c}) Top-scattering ($E_{\mathrm{x}}$ component) of the disk illuminated in free space (FS) by a plane wave with the electric field parallel to the disk axis; (\textbf{d}) Normalized top-scattering of the disk (blue curve) when illuminated by the light exiting from a waveguide propagating the TM mode. The yellow curve shows the normalized $|H_{\mathrm{y}}|^2$ field component at the disk center; (\textbf{e}) Near-field pattern of the norm of the magnetic field at the anapole state wavelength ($\lambda=1480$ nm), normalized to its maximum value; (\textbf{f}) Magnetic field lines at at $\lambda=1480$ nm under waveguide illumination. Details about the simulations are given in the Supplementary Information.}
      \label{fig:fig3}
\end{figure}

Similar results can be obtained for the magnetic anapole (Fig. \ref{fig:fig3}), though now we consider $r={450}$ nm and, in the illumination, the incident field has the main component of the magnetic field pointing along the $y$-axis. It can be seen that the magnetic anapole condition - this is, $M+T_{M}=0$ - is satisfied at a wavelength of $\SI{1470}{nm}$ for both normal (Fig. \ref{fig:fig3}(a)) and lateral (Fig. \ref{fig:fig3}(b)) illumination. 
Indeed, in comparison to the case of the electric anapole, we can see a quite broad region with reduced scattering under lateral illumination. We also performed FDTD simulations and calculated the scattered $E_{y}$ field for both free-space (Fig. \ref{fig:fig3}(c)) and waveguide (Fig. \ref{fig:fig3}(d)) illumination. In both situations, there is a dip in the top scattering response in the wavelength region where the magnetic anapole takes place. This confirms that a reduction of the out-of-plane scattering for the $x$-polarized far-field can be considered a signature of the existence of a magnetic anapole. Finally,  Figs. \ref{fig:fig3}(e) and (f) depict the magnetic field on the disk plane at the magnetic anapole wavelength ($\lambda=1480$ nm) under free-space and waveguide lateral illumination, respectively. The in-plane magnetic field in the disk shows three maxima, two nodes and two vortices of the in-plane magnetic field, as expected for a magnetic anapole \cite{Luk2017}.

\subsection{Far-field scattering measurements of the on-chip silicon disk}

To confirm our numerical predictions, we fabricated different samples containing sets of waveguide-disk circuits on a silicon-on-insulator wafer and using standard fabrication tools (see Supplementary Information). Figure \ref{fig:fig1}(b) shows a scanning electron microscope (SEM) image of one of the fabricated circuits, containing an input waveguide, a Y-junction to split up the incoming guiding signals into two paths and two waveguide terminations, with only one of them having an adjacent disk (see details in Fig. \ref{fig:fig1}(c)). This circuit allows us to measure simultaneously the top far-field scattering with and without disk for normalization purposes using the experimental set-up  described in Ref. \cite{Diaz2021}. The circuit-setup pair also allows us to inject either the TE or the TM mode into the input waveguide (see details in the Supplementary Information). In addition, using a polarizer we can filter the polarization of the scattered field to select either the $E_{\mathrm{y}}$ or the $E_{\mathrm{x}}$ component that allows to identify the scattering produced by the transverse electric or magnetic dipoles.

\begin{figure}[htbp]
      \centering
      \includegraphics[width=0.8\textwidth]{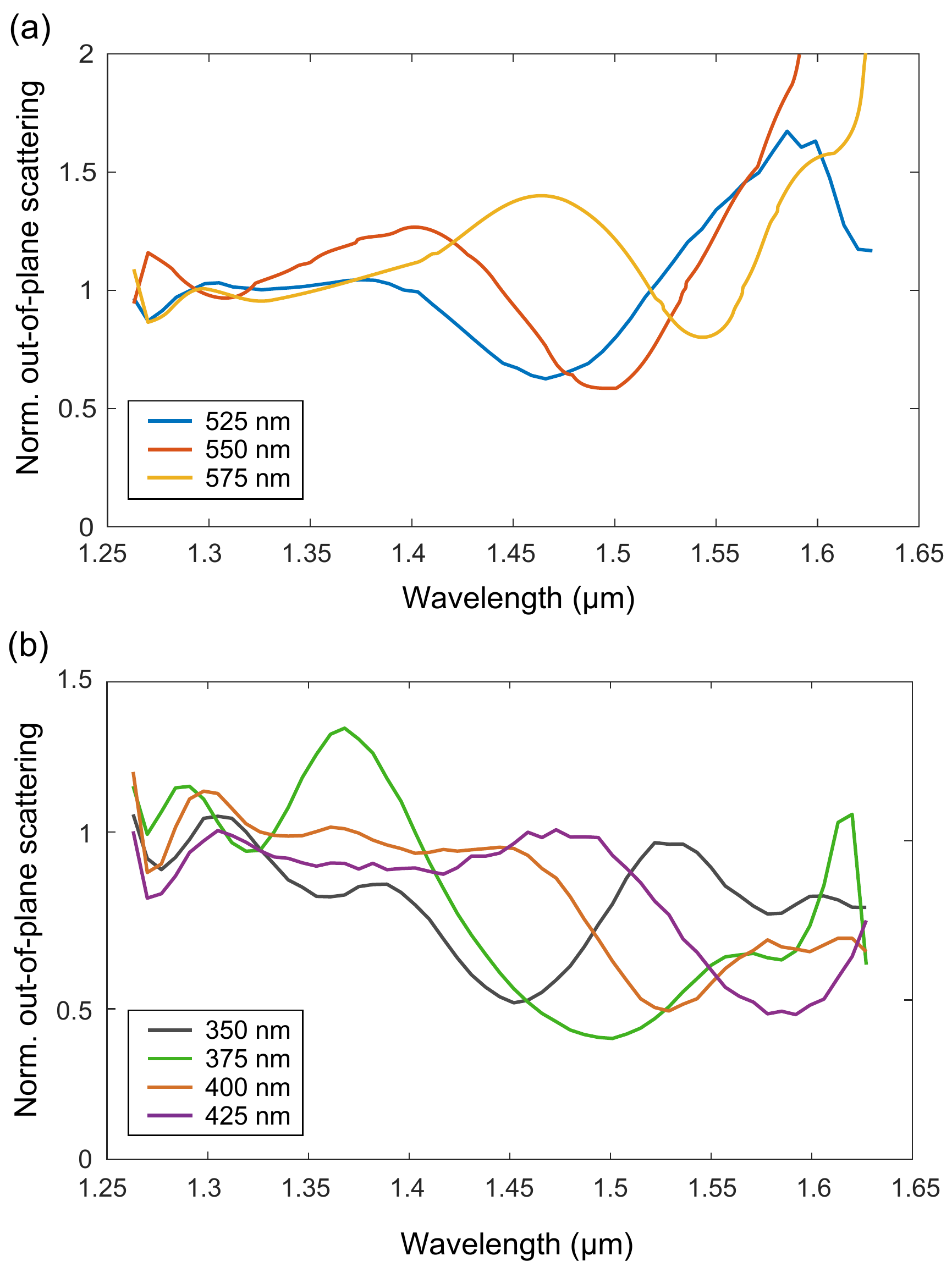}
      \caption{Experimental results of the out-of-plane scattering normalized to the value for a bare waveguide termination (no disk), recorded for disks with different radii (shown in the panels) under illumination with the TE (a) and TM (b) waveguide modes, detecting the $E_{\mathrm{y}}$ and $E_{\mathrm{x}}$ components of the far-field.}
      \label{fig:fig4}
\end{figure}

First, we characterized the second electric anapole by using the TE mode of the input waveguide. Figure \ref{fig:fig4}(a) shows the measured normalized out-of-plane scattering for three different disks having nominal radii of $525$, $550$ and $\SI{575}{nm}$. We observe that there is a wavelength region with reduced scattering (reaching even 0.5 for $r=\SI{550}{nm}$) which confirm the results of the numerical simulations presented in Fig. \ref{fig:fig2}. Notice that the overall response is slightly red-shifted as a result of the presence of the silica substrate. The top scattering is, indeed, smaller than for the case of the terminated waveguide without disk (in other words, normalized scattering $< 1$ in agreement with numerical simulations), which means that not only the scattering is well suppressed, but also that the excitation of the electric and toroidal dipoles corrects the divergence of the beam exiting the waveguide termination and contribute to reduce the overall scattering out of the chip. This was also observed in the case of the first electric anapole \cite{Diaz2021} and also occurs in the case of the magnetic anapole shown below. As expected, the wavelengths at which the scattering minimum is measured increase in proportion to the disk radius \cite{Wang2016}. Although the second electric anapole is supposed to provide a narrower response than the first one \cite{Yang2018}, we observed similar bandwidth in the scattering response for both the first \cite{Diaz2021} and second anapoles. 

Finally, we characterized the magnetic anapole by using the TM mode of the input waveguide, being the results depicted in Fig. \ref{fig:fig4}(b). Again, we observe well defined minima (reaching normalized scattering values below even 0.5) of the top scattering at wavelengths that depend on the disk radius. Again, the minima are red-shifted in comparison to the numerical simulations, which we attribute to the silica substrate.

\section{Conclusion}

We have shown that the use of integrated waveguides for on-chip excitation of wavelength-sized silicon disks enables the observation of higher-order anapoles under in-plane illumination. In particular, for the second-order electric and the magnetic anapole, a strong reduction in the top scattering is predicted from numerical simulations and observed in experimental measurements. Completing the results in Ref. \cite{Diaz2021}, we observe some differences between normal and in-plane excitation. Remarkably, for the second-order electric anapole the condition $P+T_{\mathrm{P}}=0$ is not satisfied under in-plane illumination, which can also be attributed to retardation effects in the disk. Still, our results show that the out-of-plane scattering is severely reduced, which is an important feature in the practical use of anapole states in on-chip integrated photonics. The use of the TM-like mode enables the observation of the magnetic anapole. Further engineering of the disk could lead to hybrid anapoles, where the far-field scattering arising from the spherical electric and magnetic dipoles is simultaneously cancelled \cite{Luk2017}. Our results highlight the utility of this approach (on-chip excitation) for driving complex photonic states. Indeed, by mixing TE and TM modes in the waveguide, as well as, changing the position between the waveguide termination and the disk, it could become possible to excite other complex states based on mode interference\cite{Rybin2017,Koshelev2020} for on-chip photonics.

\section{Author contributions}

A.M. conceived the idea. E.D.E., A.M., and A.B. performed the numerical simulations. A.G. fabricated the sample. All authors extensively discussed the results and contributed to the manuscript.

\section{Funding}

E.D.E. acknowledges funding from Generalitat Valenciana under grant GRISOLIAP/2018/164. A.B. acknowledges financial support by the Deutsche Forschungsgemeinschaft (DFG, German Research Foundation) through the International Research Training Group (IRTG) 2675 "Meta-ACTIVE",  project number 437527638. A.M. thanks funding from Generalitat Valenciana (Grants No. PROMETEO/2019/123, IDIFEDER/2020/041 and IDIFEDER/2021/061).  

\section{Notes} 

The authors declare no competing financial interest.

\bibliography{bibliography}

\end{document}